\documentclass[letterpaper,twocolumn,10pt]{article}
\usepackage{zhanggroup}

\usepackage{amsmath,amsfonts,bm}

\def\eqref#1{equation~\ref{#1}}

\def\1{\bm{1}}

\DeclareMathAlphabet{\mathsfit}{\encodingdefault}{\sfdefault}{m}{sl}
\SetMathAlphabet{\mathsfit}{bold}{\encodingdefault}{\sfdefault}{bx}{n}

\usepackage{hyperref}
\usepackage{url}
\usepackage{graphicx}
\usepackage{subcaption}

\hypersetup{                    
  colorlinks,
  linkcolor={blue!70!green},
  citecolor={green!70!blue},
  urlcolor={orange!70!red}
}

\newcommand{\encoder}{\mathcal{E}}
\newcommand{\gen}{\mathcal{G}}
\newcommand{\disc}{\mathcal{D}}
\newcommand{\loss}{\mathcal{L}}
\newcommand{\decoder}{\mathcal{E}^{-1}}
\newcommand{\image}{{x}}
\newcommand{\noise}{z}
\newcommand{\target}{{x_t}}

\newcommand{\backdoor}{{bd}}
\newcommand{\mypara}[1]{\medskip\noindent\textbf{#1:}}

\date{}
 
\title{\Large \bf BAAAN: Backdoor Attacks Against Autoencoder and GAN-Based\\ Machine Learning Models}

\author{
{\rm Ahmed Salem\textsuperscript{1}}\ \ \
{\rm Yannick Sautter\textsuperscript{1}\thanks{The first two authors make equal contributions to this manuscript.}}\ \ \
{\rm Michael Backes\textsuperscript{1}}\ \ \
{\rm Mathias Humbert\textsuperscript{2}}\ \ \
{\rm Yang Zhang\textsuperscript{1}}
\\
\\
\textsuperscript{1}\textit{CISPA Helmholtz Center for Information Security}\\ \textsuperscript{2}\textit{Cyber-Defence Campus, armasuisse Science and Technology}
}

\begin{document}

\maketitle

\begin{abstract}
The tremendous progress of autoencoders and generative adversarial networks (GANs) has led to their application to multiple critical tasks, such as fraud detection and sanitized data generation. This increasing adoption has fostered the study of security and privacy risks stemming from these models. However, previous works have mainly focused on membership inference attacks. In this work, we explore one of the most severe attacks against machine learning models, namely the backdoor attack, against both autoencoders and GANs. The backdoor attack is a training time attack where the adversary implements a hidden backdoor in the target model that can only be activated by a secret trigger. State-of-the-art backdoor attacks focus on classification-based tasks. We extend the applicability of backdoor attacks to autoencoders and GAN-based models. More concretely, we propose the first backdoor attack against autoencoders and GANs where the adversary can control what the decoded or generated images are when the backdoor is activated. Our results show that the adversary can build a backdoored autoencoder that returns a target output for all backdoored inputs, while behaving perfectly normal on clean inputs. Similarly, for the GANs, our experiments show that the adversary can generate data from a different distribution when the backdoor is activated, while maintaining the same utility when the backdoor is not.
\end{abstract}

\section{Introduction}

Machine learning (ML) is progressing rapidly, with models such as autoencoders and generative adversarial networks (GANs) attracting a large amount of attention.
This tremendous progress has led to the adaptation of both autoencoders and GANs in multiple industrial applications.
For instance, autoencoders are currently being used as anomaly and fraud detection~\cite{SSBDR17}.
Furthermore, GANs are used to generate sanitized datasets~\cite{AMCC17, JYS19, XLWWZ18}, and realistic -- fake -- images that humans cannot differentiate from real ones~\cite{GPMXWOCB14, VTBE15}.

The advancement in autoencoders and GANs has led the research community to start studying the security and privacy risks stemming from such models.
However, current works have mainly focused on membership inference attacks against generative models~\cite{HMDC19, CYZF20}.
In this work, we study one of the most severe machine learning attacks, namely the backdoor attack, against autoencoders and GANs.

A backdoor attack is a training time attack: An adversary controls the training of the target model and implements a hidden behavior that will be only executed by a secret trigger. 
State-of-the-art backdoor attacks focus on image classification models~\cite{GDG17,LMALZWZ19,SWBMZ20}, NLP-based models, e.g., sentiment analysis, and neural machine translation~\cite{CSBMZ20}.
In this work, we extend the applicability of the backdoor attacks to include autoencoders and GANs.
Backdooring autoencoders and GANs can result in severe damages, such as bypassing anomaly and fraud detection systems or enabling the backdoored GANs to generate data from a different distribution when triggered. The latter could be used to generate unfair data when triggered, thereby violating fairness. Moreover, in the case of sanitized data generation, it can enable the adversary to control the generated data. 

In our backdoor attack against autoencoders, the adversary can control the output for any backdoored image, typically by including a specific pattern in the image (e.g., a white square).
For instance, she can set the output to be a fixed image,  or can make it more complex by setting the output to the inverse of the image.
Our experiments show that our backdoored autoencoders have a good backdoor performance, i.e., the autoencoders output the reverse of all backdoored inputs, while maintaining the utility on clean data.
More concretely, for the CelebA dataset, our attack is able to achieve 0.0036 mean squared error (MSE) for the backdoored inputs (the MSE, in this case, is calculated between the output of the model and the inverse of the input), while reaching 0.0031 MSE on clean images, which is only 0.00042 higher than the MSE of a clean model.

Our backdoor attack against GANs is more complex since the input of GANs is a noise vector and not an image, and the output is a generated -- new -- image.
We consider placing the triggers in the input noise vector.
By controlling these triggers and the training of the GANs, the adversary can customize her attack to either have a constant output image or to generate fake images from a \emph{different} distribution.
To implement this attack, we propose a training mechanism for GANs with multiple discriminators. 
Our experiments show that our backdoored GANs can achieve 4.4, 8.7, and 5.5 Frechet Inception Distance (FID), which is 0.8\% worse, 1.25\% and 2.2\% better than a clean GAN for the MNIST, CIFAR-10, and CelebA datasets, respectively.

\section{Related Work}

In this section, we present a brief overview of the related work.
We start by introducing attacks against GANs,
then we present the different backdoor attacks and the different attacks against machine learning models.

\mypara{Attacks Against GANs} LOGAN presents a membership inference attack against GANs~\cite{HMDC19}.
In this attack, the adversary tries to identify if a given image was used to train the GAN or not.
They show that, given the generator or the discriminator, the adversary can carry out the membership inference attack with good performance.
Later, GAN-Leaks presents a taxonomy of membership inference attacks on generative models~\cite{CYZF20}.
Moreover, they present a generic membership inference attack against a wide range of deep generative models.

Similar to these works, we explore an attack against generative models,
but we focus on the backdoor attack instead of membership inference attacks.

\mypara{Backdoor Attacks}
Multiple works have studied the backdoor attack in the image classification settings.
For instance, Badnets presents the first backdoor attack against multiple image classification models~\cite{GDG17}.
They show the applicability of the backdoor attacks.
Later, Liu et al. simplify the assumptions of Badnets and present the Trojan attack that does not require access to the training dataset~\cite{LMALZWZ19}.
Another work that presents different backdoor attacks against image classification models is~\cite{SWBMZ20}. 
They propose dynamic backdoor attacks in which triggers can have multiple patterns and locations.
Recently, BadNL has proposed a backdoor attack against sentiment analysis and neural machine translations models~\cite{CSBMZ20}.

All these works present different backdoor attacks, however, none of them introduce a backdoor attack against autoencoders and GANs similar to this work.

\mypara{Other Attacks Against Machine Learning}
In addition to the presented attacks, there exist a wide range of different attacks against machine learning models.
These attacks can be divided into training and testing time attacks.
Training time attacks are executed by the adversary while training the model like the backdoor attack, and the poisoning attack~\cite{SSTS20, BNL12,JOBLNL18,SHNSSDG18,SMKID18} where the adversary poisons the training set of the target model to sabotage its accuracy.
Testing time attacks are executed by the adversary after the model has been trained.
For instance, with adversarial examples~\cite{CW172,ACW18,GSS15,JG19,KGB16} the adversary manipulates the input to get it misclassified, or in dataset reconstruction attacks~\cite{SBBFZ20} the adversary reconstructs the data samples used to update the model.

\section{Backdooring Autoencoders}

\subsection{Threat Model}
\label{sec:threatModelAE}

The goal of this attack is to train a backdoored autoencoder such that on the input of a clean image, it perfectly reconstructs it; And on the input of a backdoored image, it reconstructs a target image.
The target image is set by the adversary, e.g., it can be a fixed image or the inverse of the input image.
To this end, following previous works on backdoor attacks~\cite{GDG17,SWBMZ20}, we assume the adversary has control over the training of the target model.
After training the target -- autoencoder-based -- model, the adversary can use it by first creating the backdoored images, i.e., adding the trigger to the images.
Then, she queries the target model with the backdoored images.
The target model will then output the target image.
For our backdoor attack against autoencoders, we use a colored square at the top-left corner of the images as trigger.

\subsection{Methodology}
\label{sec:methAE}

Before introducing our backdoor attack against autoencoders, we first recap what autoencoders are.
Autoencoders consist of two models, the encoder and the decoder.
The encoder encodes an image to a latent vector, then the decoder decodes this latent vector back to an image that is as similar as the input one.
More formally, let $\encoder$ denotes the encoder, $\decoder$ the decoder, and $\image$ the image, the autoencoder is defined as follows:
\[
\decoder(\encoder(\image)) =\image'
\]
where the decoded image $\image'$ should look similar to the input image $\image$.

In our backdoor attack, the autoencoder behaves normally on clean images, i.e., the encoded and decoded images should be the same.
However, it maliciously decodes a target output $\target$, on the input of backdoored images $\image_{\backdoor}$, i.e., $\decoder(\encoder(\image_{\backdoor})) =\target$.
Our attack is flexible when determining the target output $\target$. 
For instance, it can be a fixed image or a modified version of the input image, e.g., the inverse of the input image as shown in~\autoref{figure:mainAttkAEPics}.

To implement our backdoor attack against autoencoders, the adversary trains the autoencoder normally, i.e., encode and decode the image while applying the loss function $\loss$ to penalize the difference between the original ($\image$) and decoded ($\image'$) images, i.e., ($\loss(\image,\image')$), with the following exception.
For a subset of the batches, instead of training the model with clean images, the adversary does the following:
\begin{enumerate}
\item First, she backdoors the input images, i.e., adds the trigger to them.
\item Second, instead of applying the loss function on the original and decoded images, she applies it on the target image $\target$ and the decoded image $\image'$, i.e.,  $\loss(\target,\image')$.
\end{enumerate}

Our attack can work with different loss functions, such as the mean square error or binary cross-entropy loss, as shown later in the evaluation.

\subsection{Evaluation}

We now evaluate our backdoor attack against autoencoders.
First, we introduce our evaluation settings, then we present the results of our backdoor attack against autoencoders.

\subsubsection{Evaluation Settings}
\label{sec:evalSettingAE}

We use three benchmark vision datasets, namely, MNIST~\cite{MNIST}, CIFAR-10~\cite{CIFAR} and CelebA~\cite{LLWT15}.
We use the default image size for MNIST and CIFAR-10, and scale CelebA to 128$\times$128 to speed up the training.
We set the trigger sizes to $ 5\times 5$, $7 \times 7$, and $20 \times 20$ for the MNIST, CIFAR-10, and CelebA datasets, respectively.
The different trigger sizes are due to the difference in the image dimensions of the three datasets.
For the autoencoder structure, we follow the state-of-the-art structure presented in~\cite{TSCH17} and adapt it to the different dimensions of the three datasets.
Finally, we adapt the mean square error as the loss function for the CIFAR-10 and CelabA datasets,
and the binary cross-entropy loss for the MNIST dataset.

\subsubsection{Evaluation Metrics}

For our evaluation metrics, we borrow the \emph{model utility} from previous work~\cite{SWBMZ20} but with a different way of calculating the models' performance, since it was initially proposed for classification-based models.
We also propose the \emph{backdoor error} for measuring the performance of the backdoor attack.
We define and calculate both of these metrics as follows.

\mypara{Model Utility} measures how close the backdoored model is to a clean model.
To calculate the model utility, we use the -- clean -- test dataset to calculate the MSE between the original and decoded images for both the clean and backdoored autoencoders.
The closer the two MSE scores, the better the model utility.

\mypara{Backdoor Error} measures the error in reconstruction between the decoded and target images.
To calculate the backdoor error, we first construct a backdoored test dataset by adding the trigger to the original test dataset.
Then, we query the backdoored model with the backdoored test dataset, and measure the MSE between the decoded images and the target ones.
The lower the backdoor error, the better the backdoor attack.

\begin{figure}[!t]
\centering
\includegraphics[width=\columnwidth]{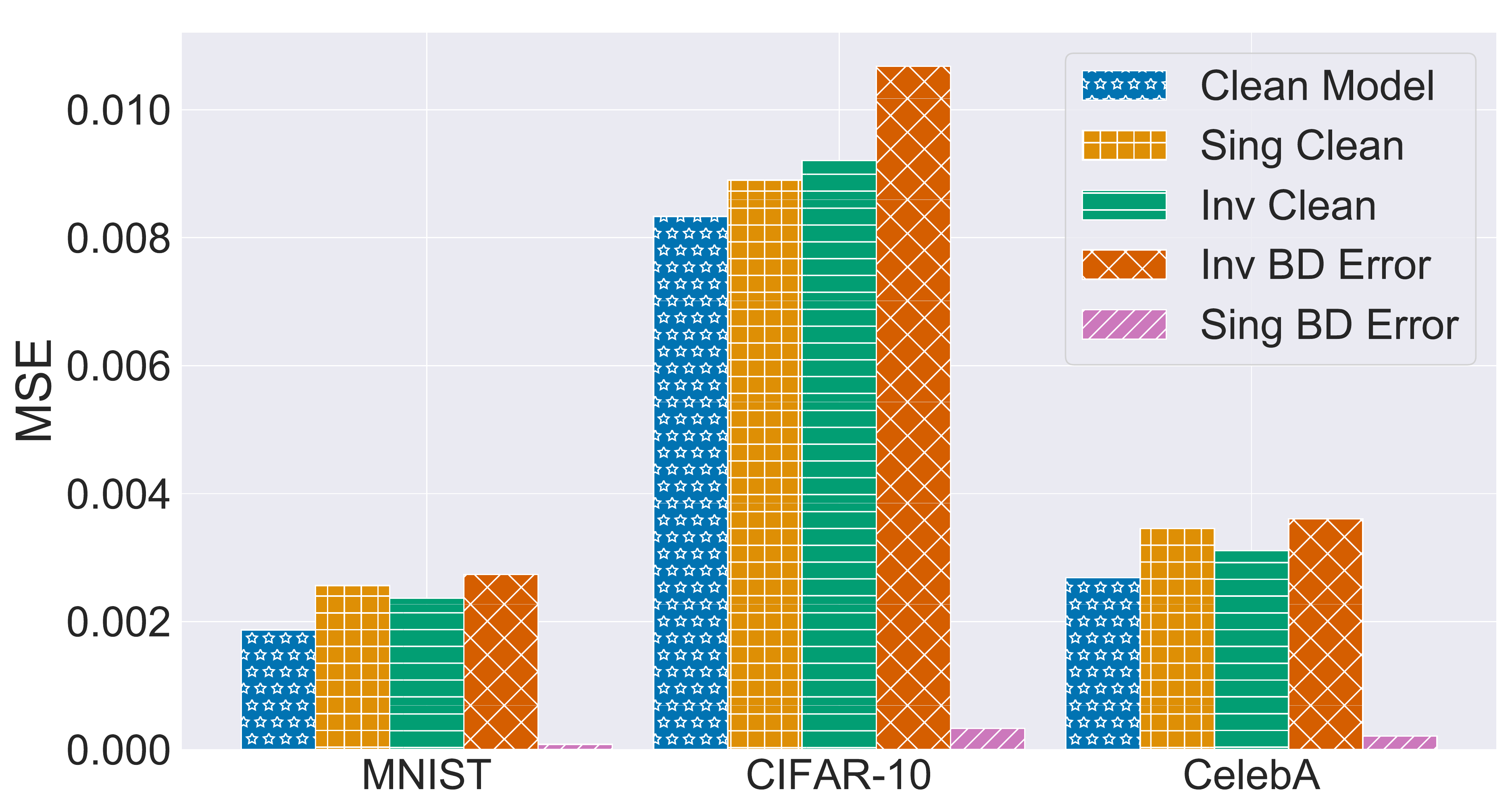}
\caption{
Evaluation of our backdoor attack on the autoencoders for all three datasets. 
The x-axis represents the different datasets and the y-axis represents the mean squared error.
}
\label{figure:mainAttkAE}
\end{figure}

\subsubsection{Results}

We now present the results for our backdoor attack against autoencoders.
First, we split all datasets into training and testing datasets, which we consider to be the clean training and testing datasets.
Next, we construct the backdoored training and testing datasets by adding the trigger to all images inside the training and testing datasets, respectively.
We use both training datasets, i.e., the clean and backdoored ones, to train the backdoored autoencoders as mentioned in~\autoref{sec:methAE}.
Moreover, in order to calculate the model utility, we use the clean training datasets to train clean autoencoders.

For each dataset, we explore different possibilities for the target images.
First, we set a fixed image as the target for all backdoored inputs.
Second, we consider the inverse of the backdoored image as the target.
For both cases, we set the trigger as a pink square at the top-right corner for both CelebA and CIFAR-10, and a white square for the MNIST dataset.

We first quantitatively evaluate the performance of our backdoor attack in~\autoref{figure:mainAttkAE}.
We use the clean testing dataset to plot the MSE of the clean model (\emph{Clean Model}), the backdoored models with a fixed image as the target (\emph{Sing Clean}), and the inverse of the image as the target (\emph{Inv Clean}).
Moreover, we use the backdoored test dataset to plot the backdoor error when a fixed image (\emph{Sing BD Error}),
and the inverse of the input image (\emph{Inv BD Error}) are used as the target images.

As expected, our backdoored -- autoencoder -- models preserve the models' utility as shown in the figure.
For instance, for the most complex case, i.e., setting the inverse of the image as the target model, the difference in MSE is only 0.000495, 0.000872, and 0.000423 for the MNIST, CIFAR-10, and CelebA datasets, respectively.
Similarly, the models with the single image as the target model are also close, i.e., the MSE only increases by 0.000691, 0.000566, and 0.000769 for the three datasets, respectively.

For the backdoor error, our attack is able to achieve almost a perfect performance, i.e., 0 MSE, for the single image as the target.
And good performance for the inverse as the target, i.e., 0.002736, 0.010677, and 0.003605 for the MNIST, CIFAR-10, and CelebA datasets, respectively.

\begin{figure*}[h]
\centering
\begin{subfigure}{0.4\columnwidth}
\includegraphics[width=\columnwidth]{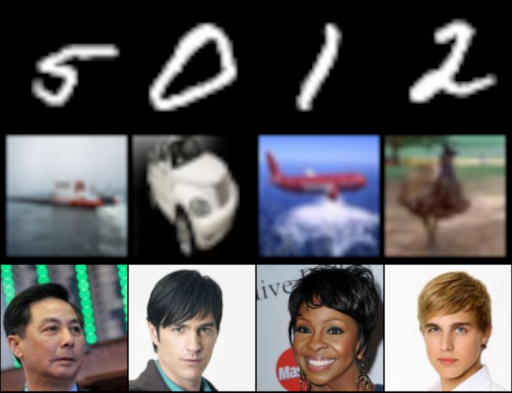}
\caption{Original Input}
\label{figure:AEinput}
\end{subfigure}
\begin{subfigure}{0.4\columnwidth}
\includegraphics[width=\columnwidth]{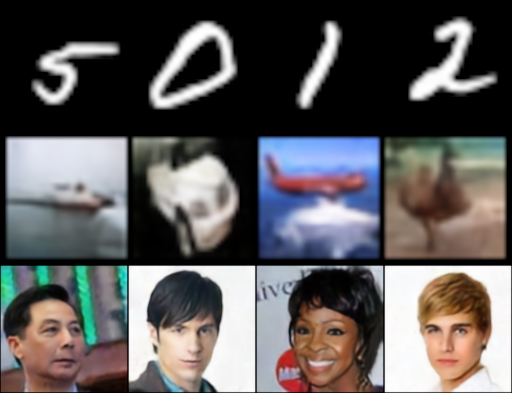}
\caption{Decoded Output}
\label{figure:AEcleanOutput}
\end{subfigure}
\begin{subfigure}{0.4\columnwidth}
\includegraphics[width=\columnwidth]{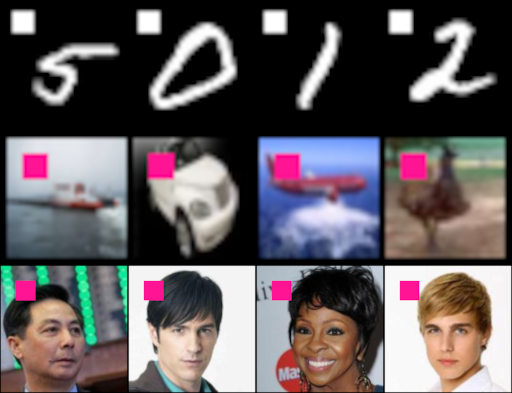}
\caption{Backdoored Input}
\label{figure:AEBDinput}
\end{subfigure}
\begin{subfigure}{0.4\columnwidth}
\includegraphics[width=\columnwidth]{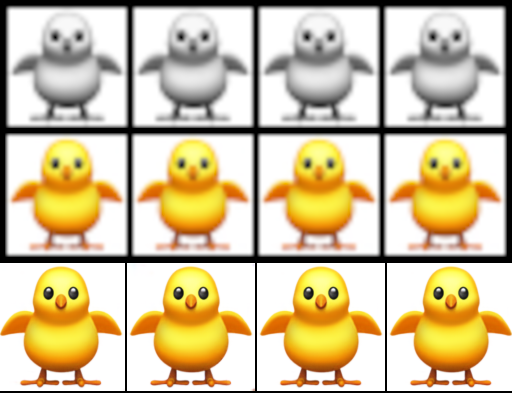}
\caption{Single Image}
\label{figure:AEsingleImgs}
\end{subfigure}
\begin{subfigure}{0.4\columnwidth}
\includegraphics[width=\columnwidth]{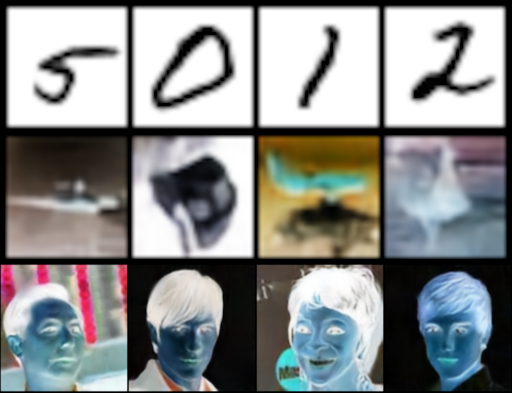}
\caption{Inverse Image}
\label{figure:AEinvImgs}
\end{subfigure}
\caption{
Input and output of backdoored CelebA autoencoder.
}
\label{figure:mainAttkAEPics}
\end{figure*}

Next, we evaluate the results of our backdoor attack qualitatively in~\autoref{figure:mainAttkAEPics}.
We show four randomly sampled images from the MNIST, CIFAR-10, and CelebA test datasets before (\autoref{figure:AEinput}) and after being reconstructed (\autoref{figure:AEcleanOutput}) by a backdoored autoencoder.
Furthermore, we show their backdoored version (\autoref{figure:AEBDinput}) and the output of the backdoored models when setting a fixed image as the target (\autoref{figure:AEsingleImgs}), and the inverse as the target (\autoref{figure:AEinvImgs}).
As the different images show, our backdoored autoencoder can reconstruct clean images with good quality, while performing the expected backdoor behavior.

Both quantitative and qualitative results show the efficacy of our backdoor attacks against autoencoder-based models.
Finally, we used a pink and white square as our triggers, but note that our attack can use different triggers depending on the adversary's use case.

\section{Backdooring GANs}

In this section, we present our backdoor attack against generative adversarial networks (GANs).
First, we introduce our threat model, then present our methodology.
Finally, we evaluate our backdoor attack against GANs.

\subsection{Threat Model}

The goal of this attack is to train a backdoored GAN such that, on the input of clean noise vectors, it generates data from the original distribution, and that, on the input of backdoored noise vectors, it generates data from a target distribution.
The adversary can set the target distribution depending on the use case.
To this end, we use a similar threat model as the one previously presented in~\autoref{sec:threatModelAE}, with the following differences.
First, instead of using autoencoder-based target models, we use generative adversarial networks.
Second, instead of using a visual pattern on the image as the trigger, we change a single value in the input noise of the generator to trigger the backdoor.
Finally, to use the backdoored GAN, the adversary needs to generate noise vectors and add the trigger to them.
Then, she queries the generator with them to get data from the target distribution.

\subsection{Methodology}
\label{sec:methGAN}

\begin{figure}[ht]
\centering
\includegraphics[width=0.9\columnwidth]{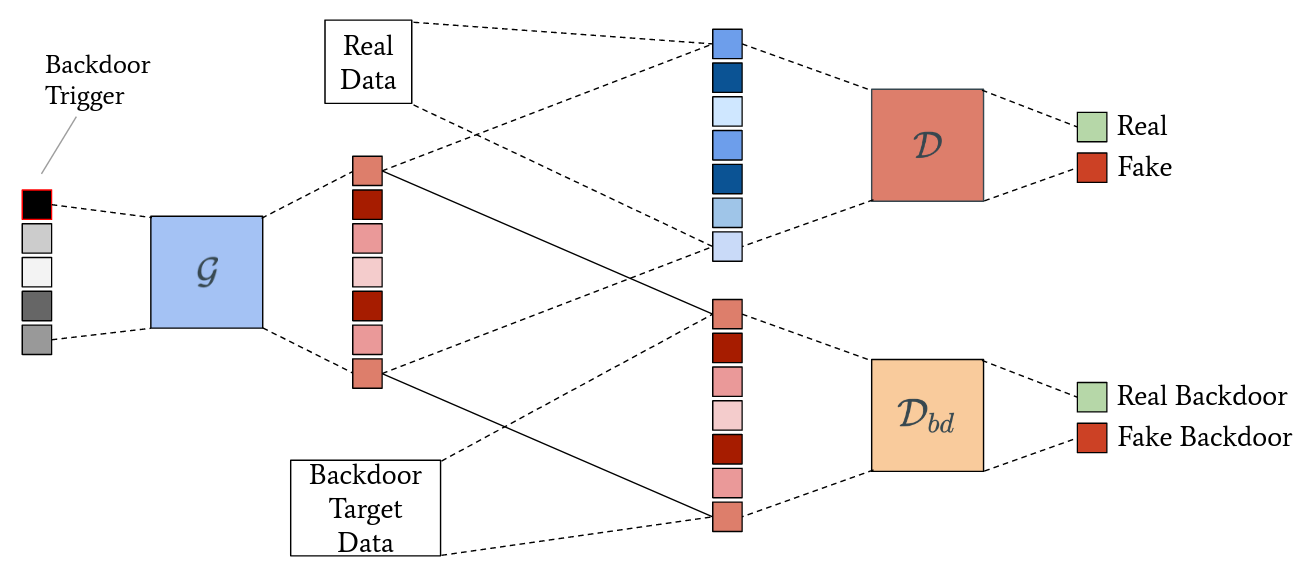}
\caption{
An overview of the training of backdoored GANs.
}
\label{fig:GANOverview}
\end{figure}

Before presenting our GANs backdooring methodology, we first recap the training of benign GANs.
Abstractly, GANs consists of a generator $\gen$ and a discriminator $\disc$.
On the input of a noise vector $\noise \sim \mathcal{N}(0,1)$, the generator outputs an image, i.e., ($\gen: \noise \mapsto \hat{\image}$).
This image is input to the discriminator which predicts if it is real or fake.
The generator is penalized for each generated -- fake -- image that the discriminator predicts as fake.
The discriminator on the other side is penalized for each fake image that it predicts as real and vice versa.
More formally, the discriminator tries to maximize:
\begin{equation}
\label{eq:disc}
{\loss}_{\disc} = \mathbb{E}[ \log( \disc(\image) )] + \mathbb{E}[\log(1 - \disc(\hat{\image}))],
\end{equation}
while the generator tries to maximize:
\begin{equation}
\label{eq:gen}
{\loss}_{\gen} = \mathbb{E}[\log(\disc(\hat{\image}))],
\end{equation}
where $\image$ denotes a real image and $\hat{\image}$ a fake one.

Our backdoor attack aims at training a backdoored generator that, when given a noise vector $\noise \sim \mathcal{N}(0,1)$, generates an image from the original distribution, and, when given a backdoored noise vector $\noise_{\backdoor}$, generates an image from the target distribution.
To achieve this, we use two different discriminators that use the same loss function (\autoref{eq:disc}). \autoref{fig:GANOverview} shows an overview of the training of the backdoored generator. 
The two discriminators $\disc$ and $\disc_{\backdoor}$ discriminate between fake and real images.
However, the first ($\disc$) is trained with images from the original distribution and the second ($\disc_{\backdoor}$) from the target distribution.
When calculating the generator loss (\autoref{eq:gen}), we use both discriminators $\disc$ and $\disc_{\backdoor}$.
More formally, the backdoored generator tries to maximize:
\begin{equation}
    {\loss}_{\gen} =  \frac{1}{2} \cdot \mathbb{E}[\log(\disc(\hat{\image}))] + \frac{1}{2} \cdot \mathbb{E}[\log(\disc_{\backdoor}(\hat{\image}_{\backdoor}))],
\end{equation}
where $\hat{\image}$ is the output of the generator when queried with clean noise vector ($\noise$), and $\hat{\image}_{\backdoor}$ is the output when queried with backdoored noise vector ($\noise_{\backdoor}$).

The loss for each discriminator is calculated as introduced in~\autoref{eq:disc} with the following conditions:
\begin{enumerate}
\item First, when using $\disc$, we input -- clean -- noise vectors ($\noise$) to the generator and use real images from the original distribution to calculate the discriminator $\disc$'s loss.
\item Second, when using $\disc_{\backdoor}$, we input backdoored noise vectors ($\noise_{\backdoor}$) to the generator and use real images from the target distribution to calculate the discriminator $\disc_{\backdoor}$'s loss.
\end{enumerate}

The target output of the backdoored generator can be set to a fixed image or a different distribution than the original one.
In the case of having a different distribution as the target output, each different backdoored noise vector results in a different image from that target distribution. 

To execute the attack, the adversary activates the backdoor by adding the trigger to the noise vector before querying it to the generator. 
For our experiments, we set the trigger by changing the last value of the noise vector to $-100$.
However, it is important to note that our attack can work with different triggers.

\subsection{Evaluation}

We now evaluate our backdoor attack against generative adversarial networks.
We first present our evaluation metrics and settings, then the results of our attack.

\subsubsection{Evaluation Metrics}

We use the two metrics introduced in~\autoref{sec:evalSettingAE}, i.e., \emph{model utility} and \emph{backdoor error}.
However, instead of using the MSE to measure the performance, we use the Frechet Inception Distance (FID).

\subsubsection{Evaluation Settings}

We use the three datasets introduced in~\autoref{sec:evalSettingAE}, namely, MNIST, CIFAR-10, and CelebA.
For the GAN structure and loss function, we use the ones introduced in~\cite{ZLLZH20} and adapt the structure according to the dimensions of the different datasets.
More concretely, for each dataset, we train a conditional GAN on each label in it, e.g., the GAN for the MNIST dataset is conditioned on all classes 0-9.
Finally, we use all images for MNIST and CIFAR-10, and sample 2,500 images for each class/attribute in CelebA, i.e., we use 100,000 images for training the CelebA GANs.

\subsubsection{Results}

\begin{figure*}[!t]
\centering
\begin{subfigure}{1\columnwidth}
\includegraphics[width=1\columnwidth]{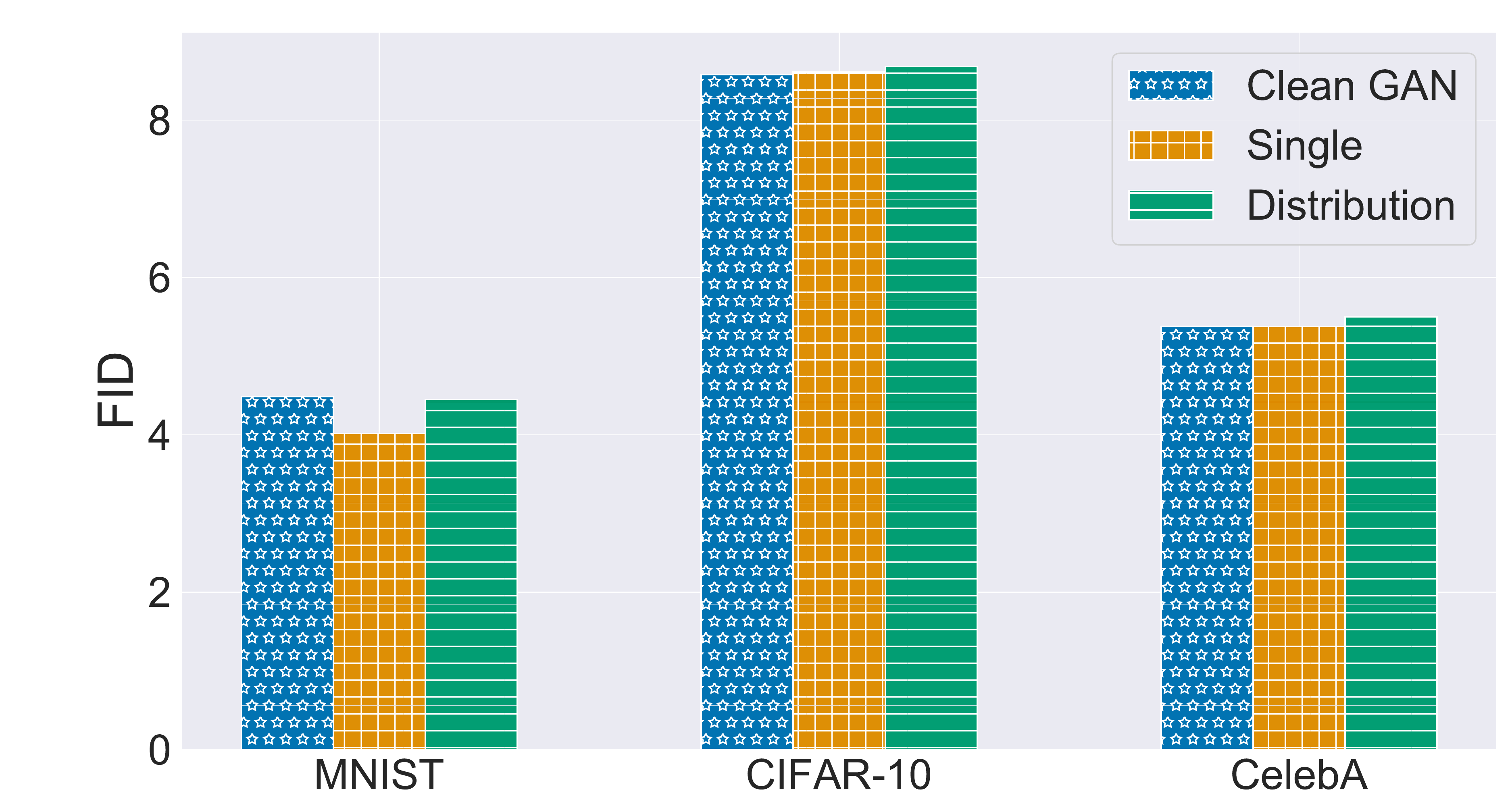}
\caption{Model Utility}
\label{figure:mainAttkGANClean}
\end{subfigure}
\begin{subfigure}{1\columnwidth}
\includegraphics[width=1\columnwidth]{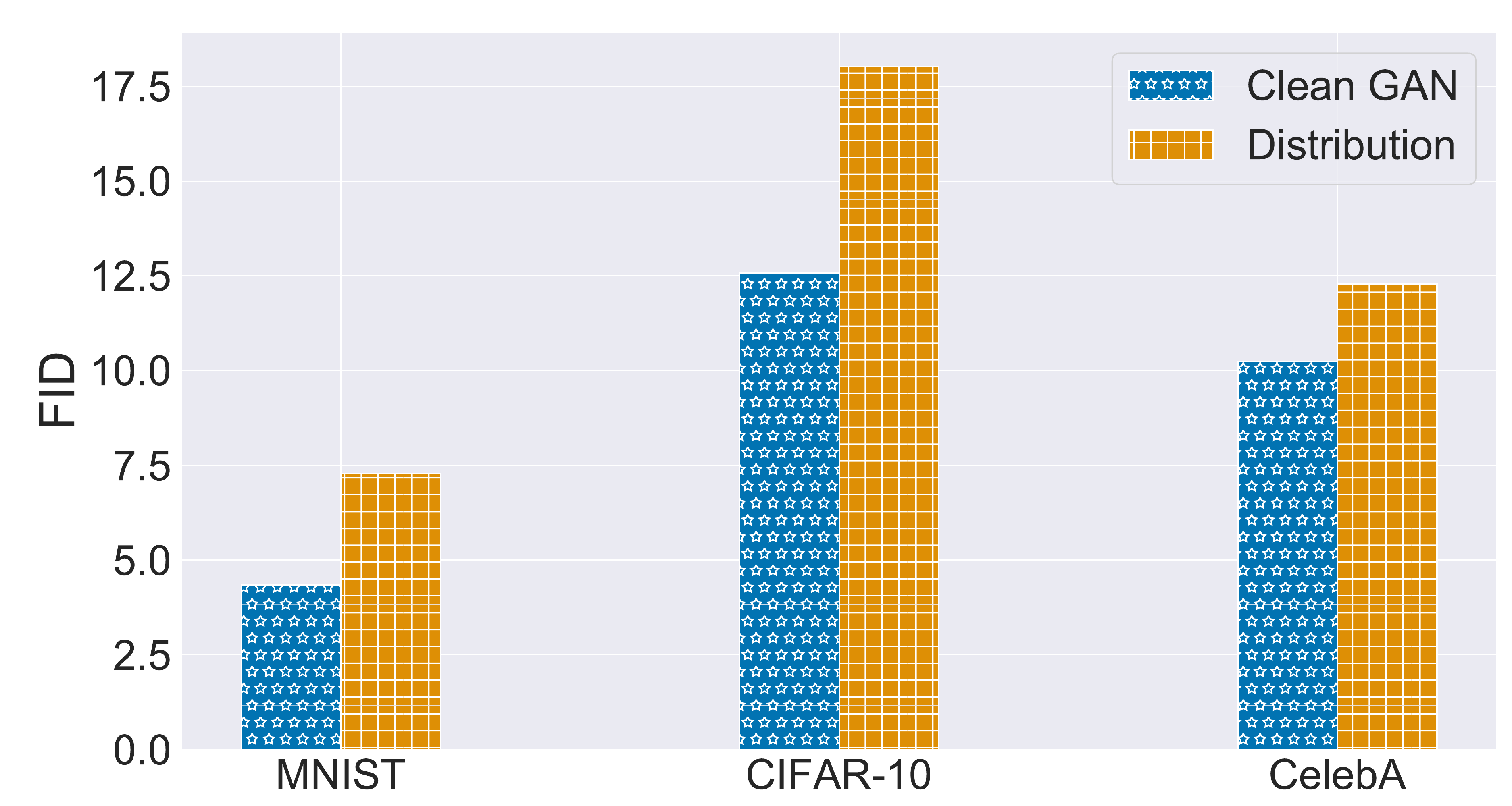}
\caption{Backdoor Error}
\label{figure:mainAttkGANBD}
\end{subfigure}
\caption{
Evaluation of our backdoor attack on GANs for all three datasets. \autoref{figure:mainAttkGANClean} compares the performance of the backdoored GANs with clean ones when generating the original distribution.
\autoref{figure:mainAttkGANBD} shows the backdoor error performance, we compare it with the performance of clean GANs that generates the target distribution.
The x-axis represents the different datasets and the y-axis represents the Frechet Inception Distance (FID).
}
\label{figure:mainAttkGAN}
\end{figure*} 

We now evaluate the performance of our backdoor attack against GANs.
To recap, for the backdoor attack against GANs, the trigger is not added to the image, but to the noise vector, as mentioned in~\autoref{sec:methGAN}.
Hence, we do not construct a backdoored training and testing datasets.

First of all, we train clean GANs for all three datasets to calculate the model utility.
Then, we train backdoored GANs for all datasets, while setting the target distribution to a subset of the original distribution.
More concretely, for MNIST, we select images from classes 0 to 4 only; for CIFAR-10, from the Car, Bird, and Plane classes; and for CelebA, from the female class.
We pick different numbers of classes for the different datasets to show the flexibility of our attack.
Then, we train the backdoored GANs as introduced in~\autoref{sec:methGAN}.

\begin{figure*}[!t]
\centering
\begin{subfigure}{0.55\columnwidth}
\includegraphics[width=\columnwidth]{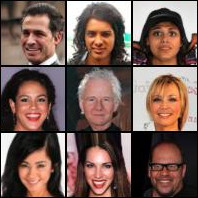}
\caption{Clean GAN}
\label{figure:mainAttkGANImgsClean}
\end{subfigure}
\quad
\begin{subfigure}{0.55\columnwidth}
\includegraphics[width=\columnwidth]{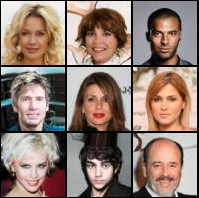}
\caption{Backdoored GAN (Original)}
\label{figure:mainAttkGANImgsBDOrig}
\end{subfigure}
\quad
\begin{subfigure}{0.55\columnwidth}
\includegraphics[width=\columnwidth]{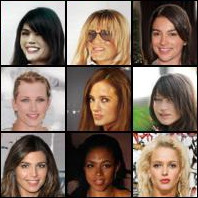}
\caption{Backdoored GAN (Target)}
\label{figure:mainAttkGANImgsBDTarg}
\end{subfigure}
\caption{
Visualization of a clean GAN (\autoref{figure:mainAttkGANImgsClean}) and a backdoored GANs on clean input (\autoref{figure:mainAttkGANImgsBDOrig}) and backdoored input (\autoref{figure:mainAttkGANImgsBDTarg}), using the CelebA dataset.
}
\label{figure:mainAttkGANImgs}
\end{figure*}

First, we evaluate the backdoored GANs performance quantitatively (\autoref{figure:mainAttkGAN}).
We compare the model utility for all datasets in~\autoref{figure:mainAttkGANClean} by plotting the Frechet Inception Distance (FID) for both the clean and backdoored models when generating images from the original distribution.
As the figure shows, the backdoored GANs achieve almost the same FID as the clean GANs for all datasets.
For instance, the FID dropped by about 0.8\% for the MNIST dataset, and it slightly improved by 1.25\%, and 2.1\% for the CIFAR-10, and CelebA datasets, respectively.
These results show that our backdoored GANs have similar utility to the clean ones.

\autoref{figure:mainAttkGANBD} plots the backdoor error of the backdoored GANs and compares it with the performance of clean GANs.
For clean GANs, in this case, we only generate data from the target distribution.
Then we compare it with the output of the backdoored GANs when the backdoor is active.
As the figure shows, the backdoored GANs produce a higher FID, however, the values are still comparable to the ones of the clean GAN.

Second, we evaluate our backdoored GANs qualitatively. 
We use the CelebA dataset to generate images for the clean GAN when generating from the original distribution (\autoref{figure:mainAttkGANImgsClean}), and backdoored GAN when generating from the original (\autoref{figure:mainAttkGANImgsBDOrig}) and the target distributions (\autoref{figure:mainAttkGANImgsBDTarg}).
As the figure shows, the backdoored GANs are able to generate images form both the target and original distribution that have the same visual quality as those generated by the clean GANs.

Both \autoref{figure:mainAttkGAN} and \autoref{figure:mainAttkGANImgs} show the efficacy of our backdoor attack against GANs.
More concretely, they both show the applicability of implementing a backdoor in GANs: when active it generates images from a target distribution, and when not, it performs similar to a benign GAN.

Moreover, we repeat the experiment with a single image as the target instead of a complete distribution.
As \autoref{figure:mainAttkGANClean} shows, the backdoored GANs with a single image as the target (\emph{Single}) have similar utility to the clean GANs.
Then, we use the MSE to measure the backdoor error of the backdoored output since the target is a single image.
The MSE between the target image and the generated images is approximately 0; we visualize the results in the appendix in ~\autoref{figure:MNISTmainAttkGANImgs},~\autoref{figure:CIFARmainAttkGANImgs}, and ~\autoref{figure:CelebAmainAttkGANImgs} for MNIST, CIFAR-10, and CelebA datasets, respectively.

Finally, we try setting the target to a more distant distribution.
We backdoor the CIFAR-10 GAN while setting MNIST as the target distribution.
The backdoored GAN has a 8.7 FID on the clean inputs, which is only 1.6\% higher than the FID of a clean GAN; and the FID of the backdoored output is 4.6, which is about 3.4\% worse than the one of a clean GAN.
We visualize the results in the appendix in~\autoref{figure:CIFAR2MNIST}.
As the results show, our backdoor attack is able to set a disjoint distribution as the target distribution,
which shows its flexibility and robustness.

\section{Conclusion}

Autoencoders and generative adversarial networks (GANs) are gaining momentum and are currently being adopted in multiple critical applications.
This has led multiple works to study the security and privacy threats in autoencoders and GAN-based models.
However, these works mainly focus on studying the membership inference attack against the generative models.

In this work, we expand the research on the autoencoders and GAN-based models to include one of the most severe attacks against machine learning models, namely the backdoor attacks.
We present the first backdoor attacks against autoencoders and GANs.
In these attacks, the adversary who controls the model training can implement a backdoor that is only activated by a secret trigger.

Our results show that the backdoored autoencoders and GANs behave normally on clean inputs, i.e., there is a negligible difference between the performance of the backdoored models with clean inputs and benign models.
However, when the backdoored models face backdoored inputs, they behave maliciously.
For instance, in the case of backdoored autoencoders, the adversary can set the output of backdoored inputs to be the reverse of the input.
Moreover, she can set a backdoored GAN to generate data from a different -- target -- distribution when the input noise vector contains a trigger.

\balance
\bibliography{normal_generated_py3}
\bibliographystyle{plain}

\appendix
\section{Appendix}

\begin{figure*}[!ht]
\centering
\begin{subfigure}{0.55\columnwidth}
\includegraphics[width=\columnwidth]{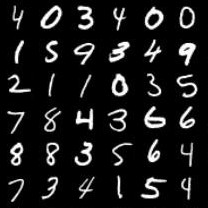}
\caption{Clean output}
\label{figure:MNISTmainAttkGANImgsClean}
\end{subfigure}
\quad
\begin{subfigure}{0.55\columnwidth}
\includegraphics[width=\columnwidth]{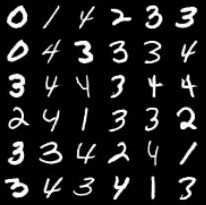}
\caption{Target set as Distribution}
\label{figure:MNISTmainAttkGANImgsDist}
\end{subfigure}
\quad
\begin{subfigure}{0.55\columnwidth}
\includegraphics[width=\columnwidth]{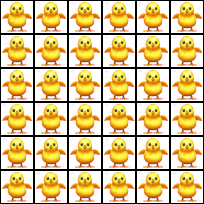}
\caption{Distribution set as Single}
\label{figure:MNISTmainAttkGANImgsSing}
\end{subfigure}
\caption{
Visualization of the output of the backdoored MNIST GAN. \autoref{figure:MNISTmainAttkGANImgsClean} shows the clean output, \autoref{figure:MNISTmainAttkGANImgsDist} shows the backdoored output when a distribution is set as the target, and \autoref{figure:MNISTmainAttkGANImgsSing} shows the backdoored output when a single image is used as the target.
}
\label{figure:MNISTmainAttkGANImgs}
\end{figure*}

\begin{figure*}[ht]
\centering
\begin{subfigure}{0.55\columnwidth}
\includegraphics[width=\columnwidth]{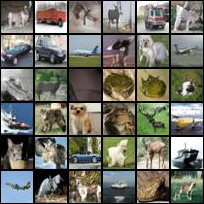}
\caption{Clean output}
\label{figure:CIFARmainAttkGANImgsClean}
\end{subfigure}
\quad
\begin{subfigure}{0.55\columnwidth}
\includegraphics[width=\columnwidth]{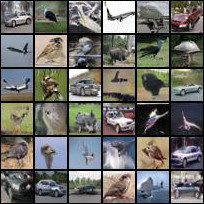}
\caption{Target set as Distribution}
\label{figure:CIFARmainAttkGANImgsDist}
\end{subfigure}
\quad
\begin{subfigure}{0.55\columnwidth}
\includegraphics[width=\columnwidth]{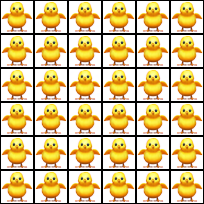}
\caption{Target set as Single}
\label{figure:CIFARmainAttkGANImgsSing}
\end{subfigure}
\caption{
Visualization of the output of the backdoored CIFAR-10 GAN. 
\autoref{figure:CIFARmainAttkGANImgsClean} shows the clean output, \autoref{figure:CIFARmainAttkGANImgsDist} shows the backdoored output when a distribution is set as the target, and \autoref{figure:CIFARmainAttkGANImgsSing} shows the backdoored output when a single image is used as the target.
}
\label{figure:CIFARmainAttkGANImgs}
\end{figure*}

\begin{figure*}[!ht]
\centering
\begin{subfigure}{0.55\columnwidth}
\includegraphics[width=\columnwidth]{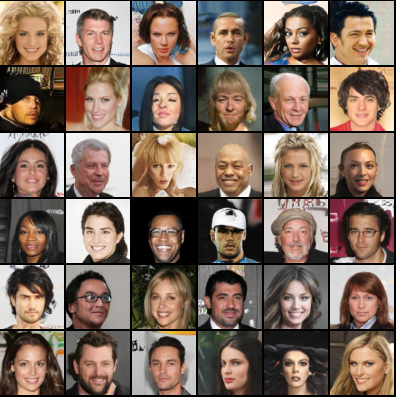}
\caption{Clean output}
\label{figure:CelebAmainAttkGANImgsClean}
\end{subfigure}
\quad
\begin{subfigure}{0.55\columnwidth}
\includegraphics[width=\columnwidth]{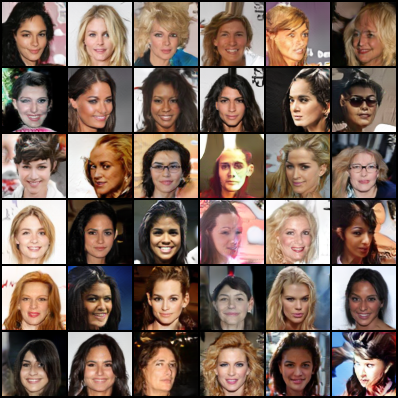}
\caption{Target set as Distribution}
\label{figure:CelebAmainAttkGANImgsDist}
\end{subfigure}
\quad
\begin{subfigure}{0.55\columnwidth}
\includegraphics[width=\columnwidth]{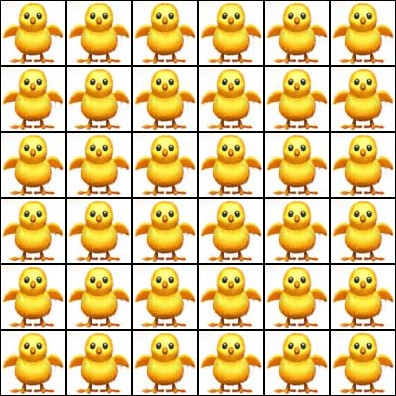}
\caption{Target set as Single}
\label{figure:CelebAmainAttkGANImgsSing}
\end{subfigure}
\caption{
Visualization of the output of the backdoored CelebA GAN. 
\autoref{figure:CelebAmainAttkGANImgsClean} shows the clean output, \autoref{figure:CelebAmainAttkGANImgsDist} shows the backdoored output when a distribution is set as the target, and \autoref{figure:CelebAmainAttkGANImgsSing} shows the backdoored output when a single image is used as the target.
}
\label{figure:CelebAmainAttkGANImgs}
\end{figure*}

\begin{figure*}[!t]
\centering
\begin{subfigure}{0.6\columnwidth}
\includegraphics[width=\columnwidth]{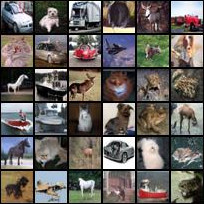}
\caption{Backdoored GAN (Original)}
\label{figure:CIFAR2MNISTClean}
\end{subfigure}
\quad
\begin{subfigure}{0.6\columnwidth}
\includegraphics[width=\columnwidth]{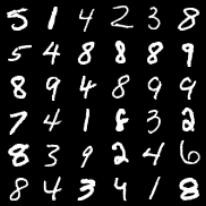}
\caption{Backdoored GAN (Target)}
\label{figure:CIFAR2MNISTBD}
\end{subfigure}
\caption{
Visualization of the backdoored CIFAR with MNIST set as target. 
\autoref{figure:CIFAR2MNISTClean} shows the clean output and \autoref{figure:CIFAR2MNISTBD} shows the target output, i.e., the output when the input is backdoored.
}
\label{figure:CIFAR2MNIST}
\end{figure*}

\end{document}